\documentclass[prl,twocolumn,superscriptaddress,amsmath,amssymb,showpacs,floatfix,preprintnumbers]{revtex4}
\usepackage{amsmath}
\usepackage{graphicx}
\usepackage{dcolumn}
\usepackage{color}
\DeclareGraphicsExtensions{.png .jpg .pdf}

\begin{document}

\title{Current-induced torques in magnetic Weyl semimetal tunnel junctions}

\author{D. J. P. de Sousa}\email{sousa020@umn.edu}
\affiliation{Department of Electrical and Computer Engineering, University of Minnesota, Minneapolis, Minnesota 55455, USA}
\author{Fei Xue}
\affiliation{Physical Measurement Laboratory, National Institute of Standards and Technology,
Gaithersburg, Maryland 20899-6202, USA}
\affiliation{Institute for Research in Electronics and Applied Physics \& Maryland Nanocenter, University of Maryland, College Park, MD 20742}
\author{J. P. Wang}
\affiliation{Department of Electrical and Computer Engineering, University of Minnesota, Minneapolis, Minnesota 55455, USA}
\author{P. M. Haney}
\affiliation{Physical Measurement Laboratory, National Institute of Standards and Technology,
Gaithersburg, Maryland 20899-6202, USA}
\author{Tony Low}\email{tlow@umn.edu}
\affiliation{Department of Electrical and Computer Engineering, University of Minnesota, Minneapolis, Minnesota 55455, USA}

\date{ \today }

\begin{abstract}
We study the current-induced torques in asymmetric magnetic tunnel junctions containing a conventional ferromagnet and a magnetic Weyl semimetal contact. The Weyl semimetal hosts chiral bulk states and topologically protected Fermi arc surface states which were found to govern the voltage behavior and efficiency of current-induced torques. We report how bulk chirality dictates the sign of the non-equilibrium torques acting on the ferromagnet and discuss the existence of large field-like torques acting on the magnetic Weyl semimetal which exceeds the theoretical maximum of conventional magnetic tunnel junctions. The latter are derived from the Fermi arc spin texture and display a counter-intuitive dependence on the Weyl nodes separation. Our results shed light on the new physics of multilayered spintronic devices comprising of magnetic Weyl semimetals, which might open doors for new energy efficient spintronic devices.

\end{abstract}


\maketitle

\emph{Introduction.} Magnetic topological materials have been a subject of intense interest in recent years~\cite{ref1, ref2, ref3, ref4, ref5, ref6}, exhibiting a unique interplay between magnetism and band structure topology~\cite{ref2, ref3, ref4, ref7}. In particular, magnetic Weyl semimetals (MWSM), three-dimensional topological semimetals with broken time-reversal symmetry, have drawn considerable attention due to their unique electronic structure~\cite{ref3, ref8, ref9}. In these materials, low-energy quasi-particles behave as chiral massless Weyl fermions~\cite{ref8}, underpinning several interesting transport phenomena such as the chiral anomaly~\cite{ref10, ref11, ref12, ref13} and chiral magnetic effect~\cite{ref14, ref15}. In addition, MWSMs also host topologically protected surface states, so-called Fermi arc (FA) states, with distinctive spin textures~\cite{ref8, ref16}.

\begin{figure}[t]
\includegraphics[width=\linewidth]{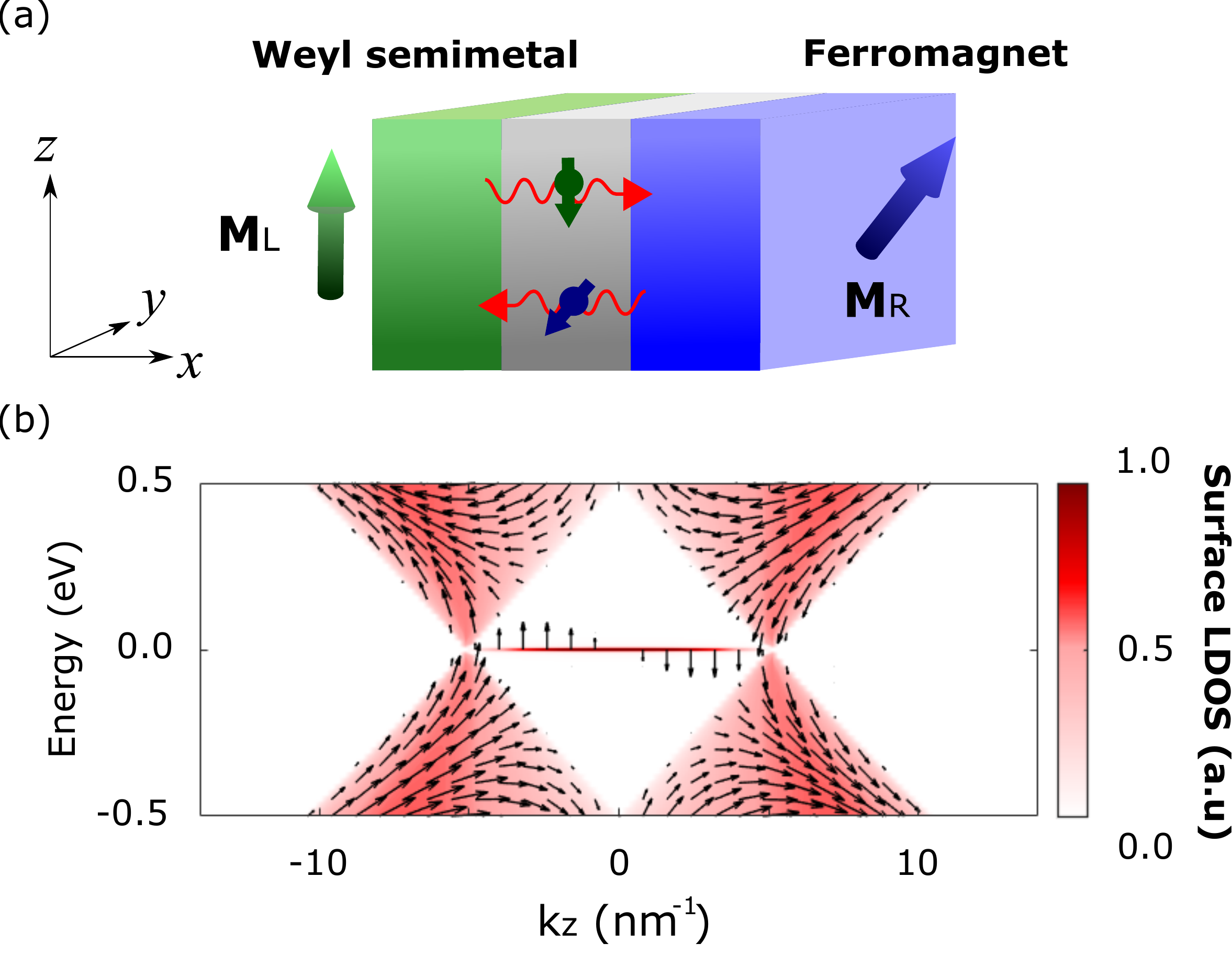}
\caption{(Color online) (a) Asymmetric MTJ comprising of a magnetic Weyl semimetal (left layer) and a trivial ferromagnetic (right layer) contact. (b) Spin-texture and local density of states at the interface of a semi-infinite magnetic Weyl semimetal system as a function of $k_z$, for $k_y=0$.  Surface normal is $\hat{\textbf{x}}$, magnetization direction is $\hat{\textbf{z}}$. The vector field is $\textbf{S} = (S_z, S_y)$ as described in Ref.~\cite{SI}. }
\label{fig1}
\end{figure}

In this letter, we show that the bulk chiral and FA surface states have a strong influence on the flow of spin-currents in magnetic tunnel junctions composed of a MWSM and a trivial ferromagnet (FM) separated by a thin insulating spacer, as illustrated in Fig.~\ref{fig1}(a). We discuss how the chirality of the MWSM bulk states dictates the characteristics of current-induced torques acting on the magnetization of the trivial FM layer, giving rise to unconventional voltage dependencies. In addition, we show that the presence of FA states at the MWSM/insulator interface naturally leads to exceptionally large field-like spin transfer torques into the MWSM layer, with a counter-intuitive dependence on the magnetic exchange interaction. Our findings highlight the novel non-equilibrium spin torques phenomena due to MWSM, offering a new perspective on the application of these systems in spintronics.

\emph{Theoretical model.} We investigate an asymmetric magnetic tunnel junctions (MTJ)~\cite{ref17, ref18, ref19} consisting of an insulating spacer sandwiched between a MWSM and a trivial FM, as sketched in Fig.~\ref{fig1}(a). We employ a 4-band tight-binding model describing a general 3D magnetic Weyl semimetal constructed on a cubic lattice of side length $a$, whose $k$-space bulk Hamiltonian is~\cite{ref20, ref21}
\begin{eqnarray}
\mathcal{H} = \tau_z \otimes [\textbf{f}(\textbf{k})\cdot \boldsymbol{\sigma}] + \tau_x \otimes [g(\textbf{k}) \sigma_0] + \tau_0 \otimes \left(\displaystyle \frac{\beta}{2}\hat{\textbf{m}}\cdot\boldsymbol{\sigma}\right),
\label{eq1}
\end{eqnarray}
where $\textbf{f}(\textbf{k}) = \hat{\textbf{x}}t\sin(k_x a)+ \hat{\textbf{y}} t\sin(k_y a)+ \hat{\textbf{z}}t\sin(k_z a)$ and $g(\textbf{k}) = t (1 - \cos(k_x a)) + t (1 - \cos(k_y a)) + t (1 - \cos(k_z a))$ are structure factors,  $\boldsymbol{\sigma} = \hat{\textbf{x}} \sigma_x + \hat{\textbf{y}} \sigma_y + \hat{\textbf{z}} \sigma_z$ is the vector of Pauli matrices and $t = 1$ eV is the nearest neighbor hopping parameter. The Pauli matrices $\boldsymbol{\tau}$ ($\boldsymbol{\sigma}$) operate in the orbital (spin) space and $\hat{\textbf{m}}$ is the unit vector pointing along the magnetization direction with $\beta$ being the exchange splitting~\cite{WeylPhase}, related to the exchange field strength $B_{\rm{exc}}$ by $B_{\rm{exc}} = \beta/2\mu_{\textrm{B}}$ where $\mu_{\textrm{B}}$ is the Bohr magneton. In constructing the MTJ structure, we discretize $\mathcal{H}$ along the transport direction and rewrite $\mathcal{H}=H_{\rm{S}} + \hat{W}_{\rm{c}} e^{i k_x x} + \hat{W}_{\rm{c}}^{\dagger} e^{-i k_x x}$, such that the MWSM is viewed as a series of principal layers (PLs) having translational invariance in the $yz$-plane, as described by $H_{\rm{S}}$, and connected to its nearest-neighbor PLs via interlayer hopping matrices $\hat{W}_{\rm{c}}$. The Hamiltonians describing the FM and insulating barrier are also constructed in a similar fashion~\cite{SI}. 

\begin{figure}[t]
\includegraphics[width=\linewidth]{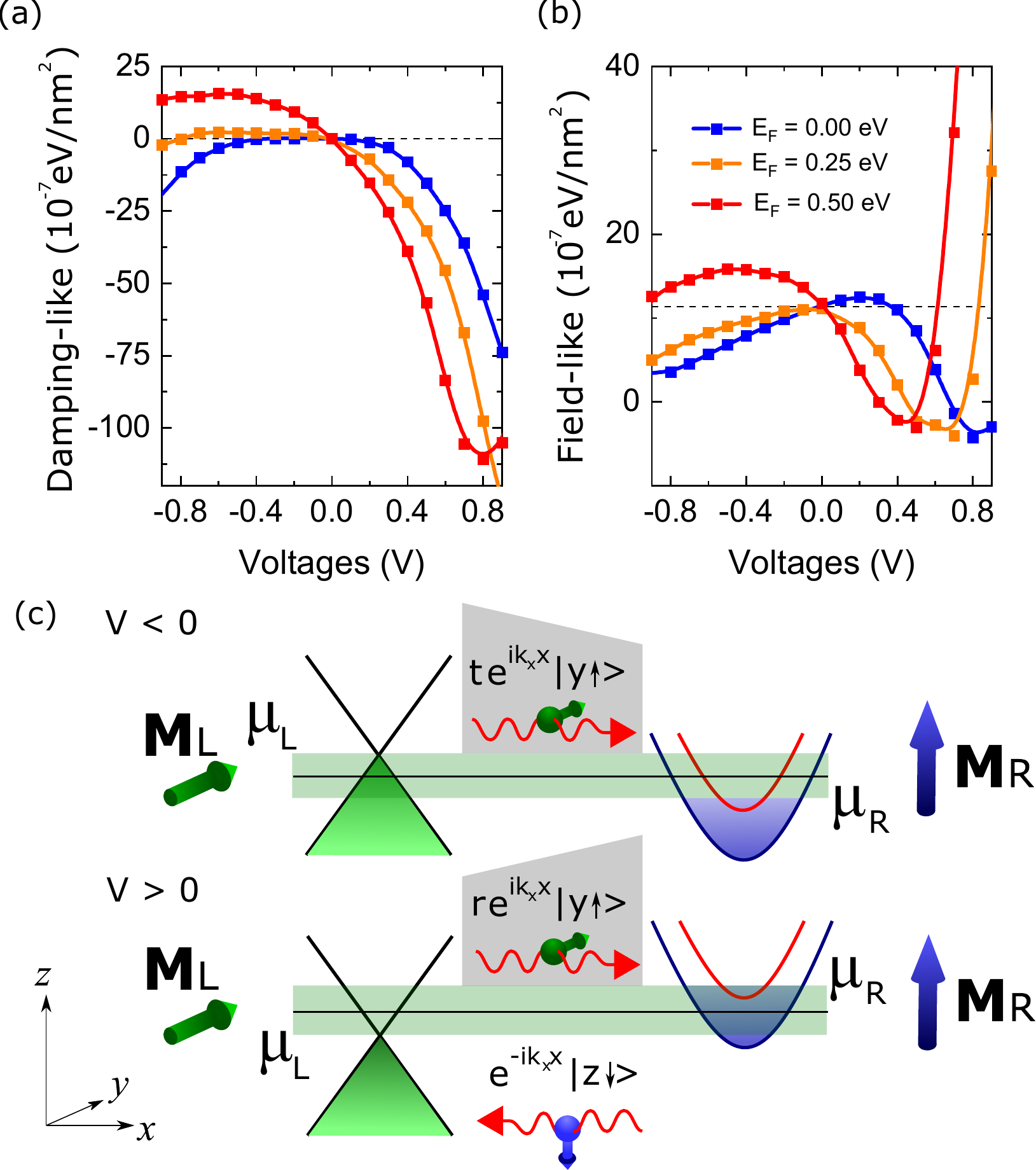}
\caption{(Color online) Panels (a) and (b) show the voltage dependence of damping-like and field-like torques, respectively, acting on the magnetization of the ferromagnetic layer at different doping levels $E_F$. The dashed horizontal line highlights the equilibrium torque values in both cases. The band diagrams are shown in panel (c) for $E_F = 0$ eV, with negative (top) and positive (bottom) voltages. Filled bands are represented by shaded regions and the spin kets represent the net spin polarization direction. The shaded region between $\mu_L$ and $\mu_R$ demarcate the transport energy window and the tunnel barrier is represented in shaded gray.}
\label{fig2}
\end{figure}

\emph{Spin texture of bulk and surface states.} We begin by reviewing the aspects of the MWSM which underlie the unique properties of the MTJ device. The key property is the chirality-derived spin texture of the bulk and surface states in $k$-space. Figure~\ref{fig1}(b) displays the surface local density of states (LDOS), $A(k_z, \epsilon)$, and associated spin texture as a function of the z component of in-plane momentum $k_z$, with $k_y = 0$, and energy $\epsilon$ as obtained from the surface Green's function of a semi-infinite MWSM lead~\cite{SI}. The surface LDOS exhibits two distinct contributions; from the FA and bulk states. The FA surface states appear as a straight line connecting the two Weyl nodes at $\textbf{k}_0^{\pm} = (0, 0, \pm k_0)$~\cite{WeylPhase}, which we describe in more detail in the next paragraph. The projected bulk states form the Dirac cones around the Weyl points, which we describe here. The origin of this behavior can be better understood by considering the long-wavelength simplified effective Hamiltonian derived from Eq.~(\ref{eq1}) for the two-crossing bands: $H = \hbar v (\sigma_{x}k_x + \sigma_{y}k_y + \kappa\sigma_{z})$ with a $k_z$-dependent mass $\kappa(k_z) = 2t (\sin(k_0a/2) - |\sin(k_za/2)|)/\hbar v$ being positive when $|k_z| < k_0$ while vanishing at $k_z = \pm k_0$. This approximation is valid to linear order in $k_{x(y)}$ while $|k_z|\leq \pi/a$ and the Fermi velocity is $v = ta/\hbar$. The expectation value of spin operator along the magnetization direction is $\langle \sigma_z \rangle = n  [\kappa(k_z)/\epsilon_{\textbf{k}}]$, where $\epsilon_{\textbf{k}} = \sqrt{k_x^2 + k_y^2 + \kappa(k_z)^2}$, with sign governed solely by the conduction ($n = +1$) and valence ($n = -1$) band indexes. The sign change can also be understood as a manifestation of the opposite electron and hole chirality at a given Weyl node~\cite{Berry}. The spin polarization is the same for both Weyl nodes at a given energy, reflecting the system magnetization as shown in Fig.~\ref{fig1}(b). The spin texture around the Weyl points results in a sign change of the spin polarization at the Weyl point energy, which is important for the behavior of the MTJ device we discuss in the next section. 

We next discuss the properties of the FA surface states, whose spin texture is shown in Fig.~\ref{fig1}(b). Because of their chiral nature, these states inherit a peculiar spin texture where, in this model, all spins are perpendicular to the surface normal and magnetization $\hat{\textbf{x}} \times \hat{\textbf{m}}$, irrespective of the exchange field strength, a feature that is not present in the bulk. For clarity, we consider the long-wavelength description of a MWSM occupying the $x < 0$ half-space (interface at $x = 0$), from which the following evanescent solutions are obtained by assuming infinite mass boundary conditions~\cite{ref23}:
\begin{eqnarray}
\Psi_{\textbf{k}}^{\textrm{FA}}(\textbf{r}) = \displaystyle C \left(
\begin{tabular}{c}
$-i$ \\
$1$
\end{tabular}\right)e^{\kappa x}e^{i k_y y} \Phi_{k_z}(z),
\label{eq3}
\end{eqnarray}
with dispersion $\epsilon_{\textbf{q}}^{\textrm{FA}} = -\hbar v k_y$, where $C$ is a normalization constant and $\Phi_{k_z}(z)$ is the periodic part of the Bloch state along the $z$ direction. The decay constant coincides with the mass term $\kappa(k_z)$, and the evanescent solution exists as long as $\kappa(k_z) \geq 0$, a condition satisfied for $-k_0\leq k_z\leq k_0$. We find a spin polarization frozen along the $y$ direction that survives deep into the bulk when $k_z \rightarrow \pm k_0$, since $\kappa^{-1} \rightarrow \infty$ in this limit. Finally, the associated density of states is energy-independent and proportional to the Weyl nodes separation $2k_0$. This indicates that as $k_0$ increases with exchange field $\beta/2\mu_{\rm{B}}$, the number of states with spin polarization perpendicular to the magnetization also increases. These salient features will result in large field-like torques acting on the MWSM with an unique dependence on the Weyl nodes separation $2k_0$ and will be discussed in more detail later.

\emph{Torques due to chiral bulk states.} We begin by considering the current-induced torque exerted on the trivial FM lead, focusing on its voltage dependence. For simplicity, we neglect spin-orbit coupling in the trivial FM lead and treat it within a minimal tight-binding approach~\cite{SI}. This enables the tracking of spin angular momentum transfer from current-carrying electrons in the trivial FM solely in terms of spin-currents. In steady state, the spin transfer torque acting on the local moments at the $i$-th PL of the trivial FM is therefore $\textbf{T}_{i} = -[\nabla \cdot \textbf{Q}]_i$, where $[\nabla \cdot \textbf{Q}]_i = \textbf{Q}_{i-1,i} - \textbf{Q}_{i,i + 1}$ in the 1D chain~\cite{ref17, ref18, ref19, SI}, and $\textbf{Q}_{i,i+1}$ is the spin current flowing between layers $i$ and $i+1$. The total spin transfer torque acting on the trivial FM is $\textbf{T} = \sum_{i \in \textrm{FM}} \textbf{T}_{i} =  \textbf{Q}_{0,1}$, corresponding to the spin current density between the last PL of the insulating spacer ($0$-th layer) and the first PL of the right magnetic lead ($1$st layer). The Hamiltonian describing the trivial FM and insulating spacer are also discretized along the transport direction~\cite{SI}. We assume that, when the system is driven out of equilibrium under an applied voltage $V = (\mu_{\rm{R}} - \mu_{\rm{L}})/e$, where $e$ is the electron charge, the potential drops linearly within the oxide spacer. Our convention is that $V < 0$ gives rise to an electron flow from the MWSM to the FM layer.

\begin{figure}[t]
\includegraphics[width=\linewidth]{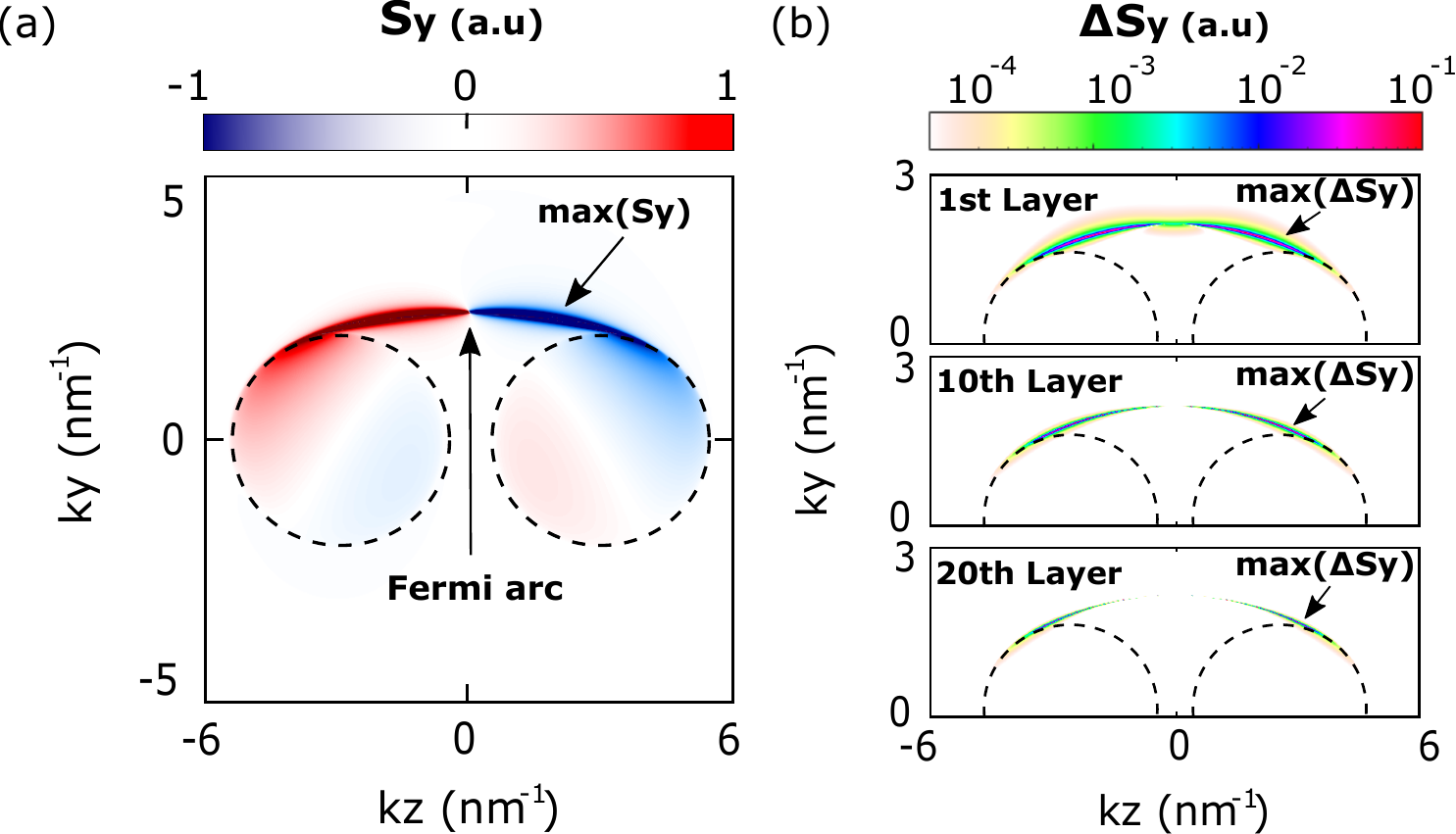}
\caption{(Color online) (a) $\textbf{k}_{||}$-resolved component of the non-equilibrium spin density $s_y(\epsilon_F, \textbf{k}_{||})$ at the first MWSM layer and (b) of the symmetric contribution, $\Delta s_y = |s_y(k_z) + s_y(-k_z)|/2$, at different MWSM layers. The dashed circles highlight the projections of bulk Fermi surfaces onto the $\textbf{k}_{||}$-space. We assume the following magnetization directions for the ferromagnet and magnetic Weyl semimetal leads $\textbf{M}_{\textrm{R}} = \hat{\textbf{y}}$ and $\textbf{M}_{\textrm{L}} = \hat{\textbf{z}}$, respectively.}
\label{fig3}
\end{figure} 

Figure~\ref{fig2} shows the voltage dependence of the current-induced damping-like and field-like torques in the FM. For simplicity, we assume $\textbf{M}_{\textrm{L}} = \hat{\textbf{y}}$, $\textbf{M}_{\textrm{R}} = \hat{\textbf{z}}$ and consider different doping levels $E_F$ in the MWSM lead, referenced from the Weyl node energy. The voltage dependence of the damping-like torque for $E_F = 0$ eV (blue symbols), is an even function of $V$ around $V = 0$. Such behavior is at odds with the well-known linear voltage dependence of damping-like torques in MTJs containing trivial FM contacts~\cite{ref24, ref25}. However, with increasing MWSM doping level, the torque curves acquires a more prominent linear voltage dependence, eventually leading to damping-like torques that act in opposite directions as we reverse the voltage polarity. This behavior can be visualized by the yellow and red symbols in Fig.~\ref{fig2}(a), for $E_F = 0.25$~eV and $E_F = 0.5$~eV, respectively.

Figure~\ref{fig2}(b) displays a similar trend for the non-equilibrium contributions of the field-like torque. The dashed horizontal line in Fig.~\ref{fig2}(b) highlights the equilibrium value, related to the equilibrium interlayer exchange coupling~\cite{ref24}. The non-equilibrium contributions are in most cases negative at small doping levels for both voltage polarities. This changes at sufficiently high doping levels, \textit{e.g.}, $E_F = 0.5~{\rm eV}$, where the non-equilibrium field-like torques act in opposite directions with the voltage polarity. 

We trace these non-trivial voltage dependencies to the opposite chiralities of bulk valence and conduction MWSM states, which in turn leads a sign change of the spin current polarization at the Weyl point energy. To show this, we have sketched the band diagrams for the case $E_F = 0$~eV in Fig.~\ref{fig2}(c) for both voltage polarities. When the system is driven out-of-equilibrium, the spin current penetrating the trivial FM layer is determined by the spin character of the right propagating states within the transport energy window $eV = \mu_{\textrm{R}} - \mu_{\textrm{L}}$, illustrated in Fig.~\ref{fig2}(c). Under negative applied voltages, only right propagating valence states of the MWSM lead can tunnel into the empty states of the FM lead while conserving energy and in-plane momentum. Hence, the states penetrating the right FM lead are $|\psi_{\textbf{t}}\rangle = te^{ik_x x}|y \uparrow\rangle $, where $|y \uparrow \rangle$ indicates the net-spin polarization direction of valence states. The spin component of the incoming state transverse to the magnetization of the trivial FM are entirely lost due to spin-dependent reflection and precession-induced dephasing of spins~\cite{ref26}, leading to a current-induced torque acting on the FM magnetization.

On the other hand, for positive applied voltages, left propagating states undergo a spin-dependent reflection, giving rise to right propagating reflected states of the form $|\psi_{\textbf{r}}\rangle = r e^{ik_x x}|y \uparrow\rangle$, since all valence states are occupied in the MWSM lead. Thus, positive and negative applied voltages lead to right-propagating states with the same net spin orientation, resulting in current-induced torques that drive the magnetization of the FM layer to the same direction. The physics is governed by the respective chirality of the bulk particle-hole states.

In contrast, if the Weyl nodes are not at half-filling, particularly in the limit where $E_F \gg |\mu_L - \mu_R|$, the energy transport window only consists of states of one chirality from the MWSM. Hence, at $V < 0$ all tunneling states would be of the form $|\psi_{\textbf{t}}\rangle = te^{i k_x x}|y\downarrow\rangle$ recovering the expected odd-in-voltage dependency of damping-like torques. At sufficiently high doping levels and large positive bias voltages $V > 0$, van Hove singularities due to additional MWSM higher energy bands within the energy transport window leads to high voltage anomalies as observed in Figs.~\ref{fig2}(a) and (b). The same argument applies to the case of large hole-doping.

\begin{figure}[t]
\includegraphics[width=\linewidth]{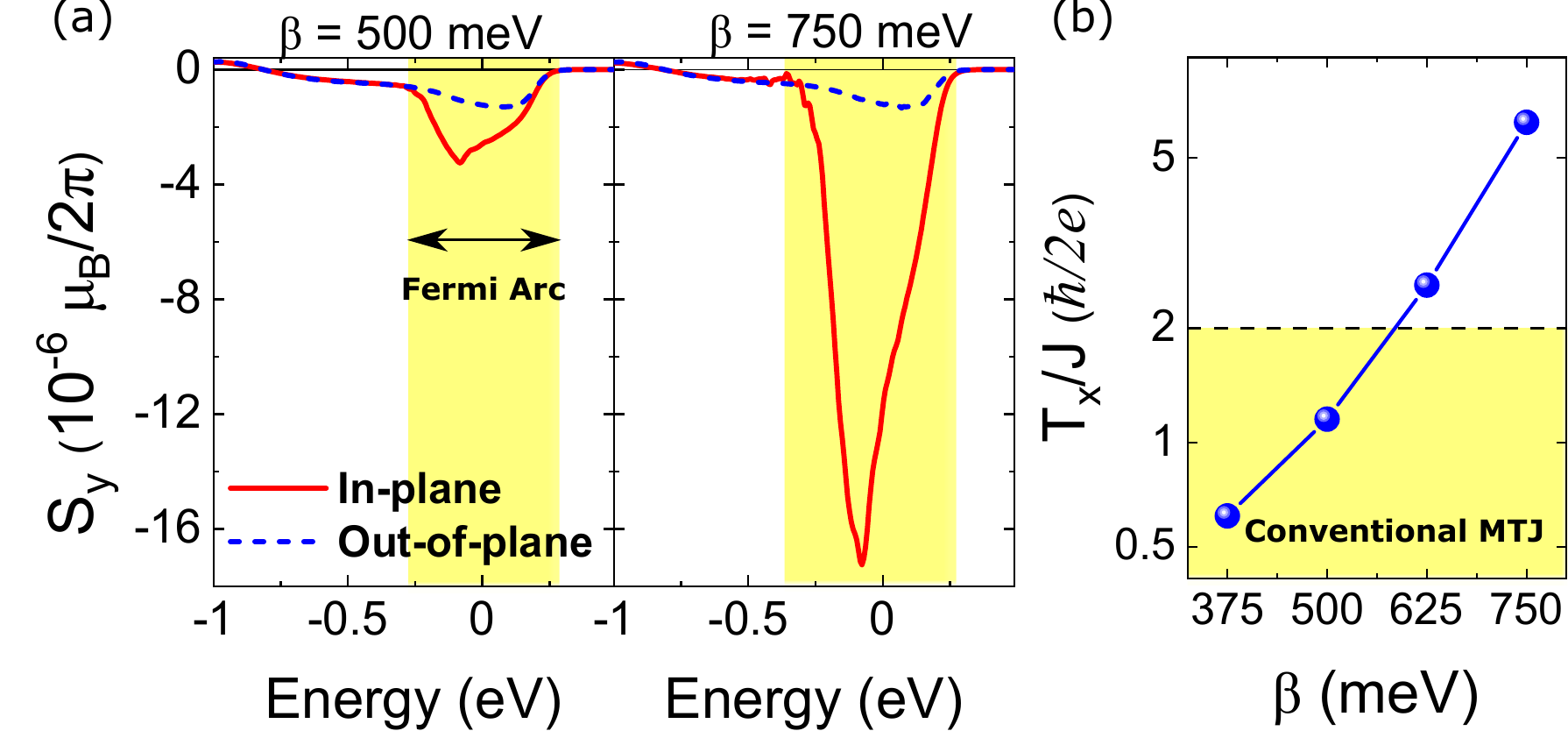}
\caption{(Color online) Energy dependence of $s_y(\epsilon) = (2 \pi)^{-2}\int d^2 \textbf{k}_{||} s_y(\epsilon, \textbf{k}_{||})$ considering $\textbf{M}_{\textrm{L}} = \hat{\textbf{z}}$ (in-plane) and $\textbf{M}_{\textrm{L}} = \hat{\textbf{x}}$ (out-of-plane) configurations for $\beta = 500$~meV (left panel) and $\beta = 750$~meV (right panel). The shaded energy window corresponds to that of Fermi arc states. (c) Field-like torque efficiency $T_x/J$ exerted on the MWSM interface. Torque efficiency in conventional MTJs is limited to the shaded region.}
\label{fig4}
\end{figure}

\emph{Torques due to Fermi arcs.} We next consider the current-induced torques exerted on the MWSM.  We find remarkably large values of field-like torque, which we ascribe to the spin texture of the Fermi arc surface states. The description of current-induced torques acting on the MWSM lead cannot be reduced to a spin-current-only calculation due to the presence of spin-orbit coupling in these systems~\cite{ref27}. Our approach is based on the computation of the layer-resolved non-equilibrium spin-density. The current-induced torque acting on the \textit{i}-th PL of the MWSM lead is $\textbf{T}_{i} = \textbf{B}_{\textrm{exc}} \times \textbf{s}_{i}$, with $\textbf{B}_{\textrm{exc}} = (\beta/2\mu_{\rm{B}})\hat{\textbf{z}}$ being the exchange field of the MWSM lead with magnetization along $\hat{\textbf{z}}$, \textit{i.e.}, $\textbf{M}_{\textrm{L}} = \hat{\textbf{z}}$. $\textbf{s}_{i}$ is the non-equilibrium spin moment density at the \textit{i}-th PL~\cite{SI}. In our calculations, we have considered different exchange field strengths for the MWSM, quantified by $\beta$, ranging from 375~meV to 750~meV~\cite{beta}, at a fixed applied voltage of $V = 0.4~{\rm V}$ so that electrons tunnel from the trivial FM to the MWSM lead. Finally, we consider $\textbf{M}_{\textrm{R}} = \hat{\textbf{y}}$ for the FM lead in the following. 

Figure~\ref{fig3}(a) displays the $\textbf{k}_{||}$-dependent non-equilibrium spin density along the $y$ direction, $s_y(\epsilon, \textbf{k}_{||})$, at the Fermi level $\epsilon = 0~{\rm eV}$ for the in-plane magnetized MWSM case. The dashed circles highlight the projections of the bulk Fermi surface onto the $\textbf{k}_{||}$-space. As can be seen, the largest contributions to $s_y(\epsilon, \textbf{k}_{||})$ comes from the FA while those from the bulk are at least $4$ orders of magnitude smaller. The total non-equilibrium spin density at a given energy $\epsilon$ is obtained by integrating $s_y(\epsilon, \textbf{k}_{||})$ over the entire Brillouin zone. However, due to the opposite chirality of states from the two Weyl points, these spin densities should exactly compensate in equilibrium. The net non-equilibrium spin density can be made apparent by considering the symmetric component $\Delta s_y = |s_y(\epsilon, k_y, k_z) + s_y(\epsilon, k_y, -k_z)|/2$. The different panels in Fig.~\ref{fig3}(b) display $\Delta s_y$ as one penetrates the MWSM lead, which clearly show that non-equilibrium spin densities decay away from the interface, but can still survive tens of layers deep into the MWSM. This indicates that FA contributions to current-induced torques are not merely confined to a few layers from the interface, as in trivial FMs~\cite{ref26}. This is due to the momentum-dependent evanescent depth of FA states which diverges at the connection points with the bulk Fermi surfaces, as discussed previously. Fig.~\ref{fig3}(b) reveals that the most relevant contributions deep into the MWSM lead come from the arc connections points with the bulk Fermi surface while the contributions around $k_z = 0$ decay very rapidly from the interface.

The interface contribution to the total spin density is shown in Fig.~\ref{fig4}(a) as a function of energy considering $\beta = 500$~meV (left panel) and $\beta = 750$~meV (right panel). We consider two distinct situations: in-plane, $\textbf{M}_{\textrm{L}} = \hat{\textbf{z}}$, and out-of-plane, $\textbf{M}_{\textrm{L}} = \hat{\textbf{x}}$, magnetized MWSM lead, while keeping $\textbf{M}_{\textrm{R}} = \hat{\textbf{y}}$ for the FM. Because the FA density of states is proportional to the in-plane projection of $\textbf{M}_{\textrm{L}}$, these two cases allow us to isolate contributions of FA states from that of bulk states. Figure~\ref{fig4}(a) shows that in-plane and out-of-plane configurations contribute the same $s_y$, except in a narrow energy window (shaded region) where there is an extra enhancement in the in-plane case. This energy window coincides with the FA states, confirming their role in the enhanced current induced field-like torques. Additional evidence of the FA role is provided by comparing left and right panels of Fig.~\ref{fig4}(a), where we show $s_y(\epsilon)$ for $\beta = 500$~meV and $\beta = 750$~meV, respectively. As is apparent, the extra contribution due to FAs increases very rapidly with $\beta$. Such behavior is counter-intuitive: one would expect that increasing the exchange field causes a less prominent contribution from those spin states misaligned with the magnetization due to spin-orbit coupling.

This discrepancy can be reconciled by noting that the density of FA states increases with $\beta$ due to the larger separation of Weyl nodes. This in turn results in an enhancement of $s_y$ within the FA energy window because its spin eigenstates are locked to the $\pm y$ direction, as previously discussed. Such feature leads to interfacial field-like torques that increase very prominently with $\beta$, as shown in Fig.~\ref{fig4}(b). In particular, the interfacial field-like torque efficiency, $T_x/J$ where $J$ is the current density, surpasses the theoretical limit of $\approx 2 (\hbar/2e)$ expected for conventional MTJs. These results indicate that field-like torques might play an important role to the magnetization dynamics of MWSM systems.

\emph{Conclusion.} 
We have studied the current-induced torques in magnetic tunnel junctions containing a magnetic Weyl semimetal contact. Our results show that the presence of magnetic Weyl semimetals substantially modifies the behavior current-induced torques. First, the chirality of electronic bulk states gives rise to anomalous voltage dependencies of spin transfer torque acting on a trivial ferromagnetic layer. Second, the presence of topologically protected Fermi arc states was found to produce giant field-like torques acting on the Weyl semimetal, in conjunction with a counter-intuitive behavior where the torque increases with exchange field strength. Most MWSMs discovered to-date consists of multiple pairs of Weyl points with crossing bands. Nevertheless, the minimal model herein allows us to elucidate on the new non-equilibrium spin torque physics, which could underpin a new generation of spintronics with energy efficient magnetization switching.

\emph{Acknowledgments.} DS, TL, and JPW were partially supported by the DARPA ERI FRANC program. TL and JPW were also partially supported by the SMART, one of two nCORE research centers.


\begin{thebibliography}{apsrev}
\bibitem{beta}
In realistic systems, the Weyl nodes separation, $2 k_0$, is the real physical parameter to be considered. In general, $k_0$ is determined by the exchange field strength and the orbital overlap with neighboring sites.
\end{thebibliography}
\end{document}